\documentclass[prd,aps,preprint,tightenlines,superscriptaddress]{revtex4}
\usepackage{epsfig}
\usepackage{axodraw}
\usepackage{amsmath}
\begin{document}

\preprint{DOE/ER/40762-332 \ UM-PP\#05-027}

\title{Factorization of Large-$x$ Quark Distributions in a Hadron}
\author{Xiangdong Ji}
\affiliation{Department of Physics, University of Maryland,
College Park, Maryland 20742, USA} \affiliation{Institute of
Theoretical Physics, Academia Sinica, Beijing, 100080, P. R.
China}
\author{Jian-Ping Ma}
\affiliation{Institute of Theoretical Physics, Academia Sinica,
Beijing, 100080, P. R. China}
\author{Feng Yuan}
\affiliation{RIKEN-BNL Research Center, Brookhaven National
Laboratory, Upton, NY 11973, USA}

\date{\today}
\vspace{0.5in}
\begin{abstract}

We present a factorization formula for valence quark distributions
in a hadron in $x\rightarrow 1$ limit. For the example of pion, we
arrive at the form of factorization by analyzing momentum flow in
the leading and high-order Feynman diagrams. The result confirms
the well-known $1-x$ scaling rule to all orders in perturbation
theory, providing the non-perturbative matrix elements for the
infrared-divergence factors. We comment on re-summation of
perturbative single and double logarithms in $1-x$.

\end{abstract}

\maketitle

Three decades ago, Brodsky, Farrar and others made a perturbative
quantum chromodyanmics (pQCD) prediction about the power behavior
of parton distributions as $x\rightarrow 1$, i.e., the density of
quarks carrying almost all the longitudinal (plus) momentum of a
hadron participating in hard scattering
\cite{Gunion:1973nm,Blankenbecler:1974tm,Farrar:1975yb,Lepage:1980fj}.
Through calculating the leading pQCD diagrams, they showed that
the valence quark distribution goes like, for example, $(1-x)^2$
in the pion and $(1-x)^3$ in the nucleon. The basic argument is
that when the valence quark carries nearly all the plus momentum,
the relevant QCD configurations in hadron wave functions must be
far off-shell and hence are amenable to pQCD treatment. The result
has since been generalized to the sea quarks, gluons,
helicity-dependent distributions \cite{Brodsky:1994kg}, and lately
to generalized parton distributions \cite{Yuan:2003fs}. The above
power law is consistent with the large-momentum behavior of the
elastic form factors through the so-called Drell-Yan-West relation
\cite{Drell:1969km,West:1976tn,Melnitchouk:2001eh}.

Despite its elegance and usefulness, a rigorous derivation of the
scaling rule in pQCD has not, to the authors' knowledge, been
established in the literature. Indeed, if one follows the original
calculations, the scaling term has a power-like infrared-divergent
coefficient $
    \int {d^2k_\perp}/{(k_\perp^2)^2} $
for the pion, and similarly for the nucleon. One might argue that
this divergence will be tamed by non-perturbative QCD effects and
hence does not affect power counting. A more satisfactory
solution, however, is to formulate a factorization theorem
explicitly separating the long distance contributions from short
distance ones. QCD factorization will establish the scaling rule
through a systematic QCD power counting, and hence validate the
result to all orders in perturbation theory. It will also provide
well-defined non-perturbative matrix elements to absorb infrared
divergences. Finally, factorization will provide a convenient tool
to re-sum possible large Sudakov double logarithms present in
$x\rightarrow 1$ limit.

The goal of this paper is to establish a factorization formula for
the valence quark distributions in a hadron in the large-$x$
limit. We consider the example of a pion, although similar studies
apply to other hadrons such as the nucleon. The factorization
arises from a detailed analysis of momentum flow in the
lowest-order pQCD diagrams and their generalization to all orders.
Without an explicit calculation, an infrared power counting allows
a straightforward determination of the $1-x$ power behavior. The
factorized expression contains hard parts, initial and final state
jets, and a soft function, as easily seen from the leading $1-x$
structure of reduced diagrams. We introduce a non-perturbative
matrix element for the soft factor which absorbs the perturbative
infrared divergence mentioned earlier.

The large-$x$ behavior of the valence quark distributions in a
pion was obtained from calculating the lowest-order Feynman
diagrams, one of which is reproduced in Fig. 1a. Let us choose the
light-cone system of coordinates defined by two light-cone vectors
$p^\mu =\Lambda (1,0,0,1)$ and $n^\mu=(1,0,0,-1)/(2\Lambda)$
($p^2=n^2=0$, $p\cdot n=1$). Any vector $k^\mu$ can be expanded in
terms of $p^\mu$ and $n^\mu$, as $k^\mu = k\cdot n p^\mu + k\cdot
p n^\mu  + k_\perp^\mu$. We choose the pion momentum mainly along
the $p$-direction ($p^+$ taken as {\it large}), $P^\mu = p^\mu +
m_\pi^2/2 n^\mu$, ignoring the $m_\pi^2$-term except when it is
needed to regularize infrared divergences. As labelled in the
figure, the incoming quark carries momentum $x_1p^\mu$ and
anti-quark momentum $ (1-x_1)p^\mu\equiv \bar x_1p^\mu $. The
anti-quark going through the final state cut (shown by the dashed
line) has momentum $k^\mu$ integrated over.

Instead of doing an explicit calculation to reproduce the familiar
result $q(x)\sim f_\pi^2(1-x)^2\int d^2k_\perp/(k_\perp^2)^2$, let
us analyze the momentum flow of the loop integral over $k^\mu$. In
the $x\rightarrow 1$ limit, the $p$ (or plus) component of $k$ is
constrained to $(1-x)p$ and is soft, $\sim \lambda$. If the
transverse momentum $k_\perp$ is on the order of $(1-x)^{1/2}p^+$,
the $n$ component of $k$ must be very large, $k^- =
k_\perp^2/2(1-x)p^+\sim p^+$. Then the anti-quark line is
jet-like, and going along the $n$-direction. The corresponding
reduced diagram is shown in Fig. 1b. If, on the other hand,
$k_\perp$ is soft $(1-x)p^+$, $k^-$ also remains soft. The
anti-quark line is now a soft line. The corresponding reduced
diagram is shown in Fig. 1c. Using an infrared power counting
generalizable to all orders in perturbation theory, one can obtain
$q(x)\sim (1-x)^2$. We refer the reader to well-known references
for the details of infrared power counting
\cite{stermanbook,CSS89}.

\begin{figure}[t]
\begin{center}
\epsfig{file=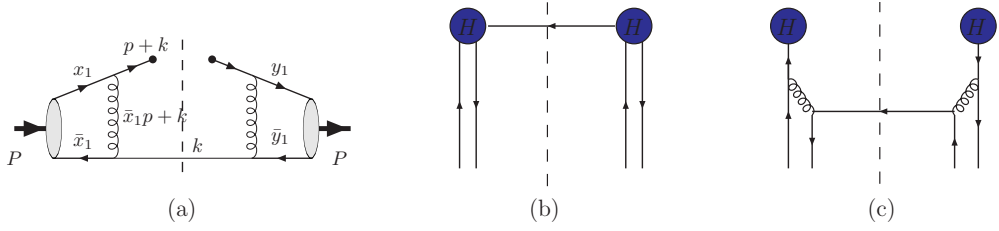,clip=,height=3.0cm,angle=0}
\end{center}
\caption{\it a). A lowest-order Feynman diagram contributing to
the leading-$1-x$ power of valence quark distribution in a pion.
b). A reduced diagram in which the final state anti-quark is a jet
along the $n$-direction. c). An alternative reduced diagram in
which the final state anti-quark is soft.}
\end{figure}

If the anti-quark is collinear to the $n$-direction (Fig.1b), the
infrared power of the diagram is $0$, obtained from the
final-state quark line as a jet, $\int
d^4k\delta(k^2)\delta(k^+-(1-x)p^+)$. The polarization sum of the
final-state quark is proportional to $k^-\sim p^+$ and hence does
not contribute any soft power. However, an explicit calculation
also finds a factor $k_\perp^2$ in the hard part. This suppression
factor is not included in ordinary power counting, and is
resulting from the spin structure of the diagram. Taking into
account this additional contribution, the actual infrared scaling
of Fig. 1b is $\lambda$. Infrared factors of a Feynman diagram
come from both $1-x$ and transverse momentum. The dimensionless
quark distribution is proportional to $f_\pi^2$ from the pion wave
function. Thus the transverse-momentum integral must be
proportional to $
    f_\pi^2 \int {d^{2} k_\perp}/{(k_\perp^2)^2} $.
Hence, the quark distribution has $(1-x)$-related infrared power
$2$, or is proportional to $(1-x)^2$. The same counting applies to
all other leading-order diagrams which are not shown explicitly.

When the final-state quark is soft, the gluon and internal quark
lines become collinear to the pion momentum (Fig.1c). The
denominators in the pion jets contribute $\lambda^{-4}$. Two
collinear-quark-gluon couplings contribute soft-power $\lambda^2$:
Usually a coupling contributed by a collinear transverse-momentum
$k_\perp$ is counted as $\lambda^{1/2}$. In the present case,
however, the transverse momentum in the pion jets originates from
the soft quark and thus must be treated as the soft scale
$\lambda$. The final-state soft quark contributes $\lambda^2$. So
the total soft power is $0$. After factoring the
transverse-momentum integral, one has a $(1-x)$-related soft-power
2, or the pion distribution goes like $(1-x)^2$, just like in the
first region, Fig. 1b.


The above consideration and result are generic. For a general
Feynman diagram of an arbitrary order contributing to parton
distributions at $x\rightarrow 1$, the momentum integrations
contain many different regions (pinched surfaces)
\cite{stermanbook,CSS89}, each of which can be represented by a
product of the various parts: the two incoming jets along the
$p$-direction ($J_{L,R}$), the out-going jets along the
$n$-direction ($J_\beta$), and two hard parts ($H_{L,R}$), each on
the different sides of the cuts connecting the $p$ and $n$ jets,
and finally a soft part connecting all jets and hard parts, as
shown in Fig. 2a. Similar factorization for the quark distribution
in a quark has been considered in
\cite{Sterman:1986aj,Korchemsky:1988si,Berger:2002sv}

\begin{figure}[t]
\begin{center}
\epsfig{file=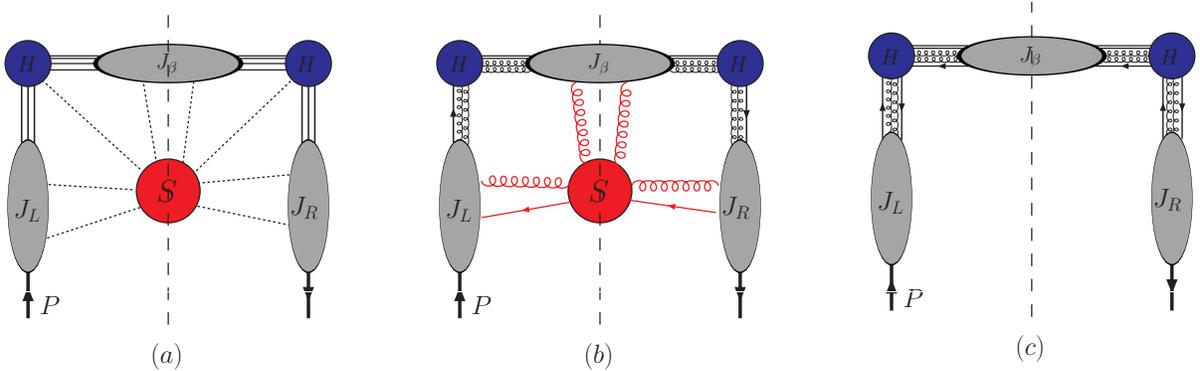,clip=,height=5.0cm,angle=0}
\end{center}
\caption{\it a). A general reduced diagram for parton
distributions at $x\rightarrow 1$. b). A leading region where the
anti-quark goes through the soft part and the final state jet
containing the eikonal line and longitudinally polarized gluons.
b). Another leading region where the anti-quark is part of the
final-state jet. There is no soft factor here because the initial
and final state jets are all color neutral.}
\end{figure}

Let us use $p_J$ to denote the number of collinear quarks or
gluons with-physical-polarization lines entering the hard part
from jet $J$; use $l_j$ to represent the number of collinear
gluons of longitudinal polarization through a similar attachment;
use $E_J^{b,f}$ to denote the number of soft bosons or fermions
connecting to the left or right hard part; use $E^{b,f}_{HL,R}$ to
label the number of soft bosons of fermions connecting to the left
or right hard part; and finally use $v_J^{3}$ to label the number
of three-point vertices in the jet, and $s_J$ the number of soft
gluons with scalar polarization and soft quarks attaching to the
jet. Then the soft part has infrared power,
\begin{eqnarray}
   \omega_S &=& E^b_{J,L}+ E^b_{J,R} + E^b_{J,\beta} + E^b_{HL} +
   E^b_{HR} \nonumber \\
    && +  \frac{3}{2}\left(E^b_{J,L}+ E^b_{J,R} + E^b_{J,\beta} + E^b_{HL} +
   E^b_{HR}\right) \ .
\end{eqnarray}
On the other hand, the infrared power associated with each jet is
\begin{equation}
    \omega_J = \frac{1}{2}(p_J-s_J -E_J^b-E^f_J-I_J) + \frac{1}{2}
      \left(s_J + l_J-v^{(3)}\right)\theta \left(s_J +
      l_J-v^{(3)}\right)\ .
\end{equation}
Combining the above results, one finds
\begin{eqnarray}
  \omega(G)&\ge&
  \frac{1}{2}\left(p_{J_{L}}+p_{J_{R}}+
  p_{J_{\beta}}\right) - 2 - \frac{1}{2}(I_{J_L} + I_{J_R}) \nonumber \\
  &&  + \frac{1}{2}\left(E^b_{J_{L}}+E^b_{J_{R}}
  +E^b_{J_{\beta}}\right) - \frac{1}{2}\left(s_{J_{L}}+s_{J_{R}}+
  s_{J_{\beta}} \right) \nonumber \\
    && + E^f_{J_{L}}+E^f_{J_{R}}+
  E^f_{\beta} + E^b_{HL}+E^b_{HR} + \frac{3}{2}(E^f_{HL}+E^f_{HR})
  \nonumber \\
  && + \frac{1}{2}\left(s_{J_{L}} + l_{J_{L}} -
  v_{J_{L}}^{(3)}\right)\theta \left(s_{J_{L}} + l_{J_{L}} -
  v_{J_{L}}^{(3)}\right) \nonumber \\
    && + \frac{1}{2}\left(s_{J_{R}} + l_{J_{R}} -
  v_{J_{R}}^{(3)}\right)\theta \left(s_{J_{R}} + l_{J_{R}} -
  v_{J_{R}}^{(3)}\right) \nonumber \\
    && + \frac{1}{2}\left(s_{J_{\beta}} + l_{J_{\beta}} -
  v_{J_{\beta}}^{(3)}\right)\theta \left(s_{J_{\beta}} + l_{J_{\beta}} -
  v_{J_{\beta}}^{(3)}\right) \ .
\end{eqnarray}
where we have taken into account the constraint that the plus
momentum going through the final state cut is $(1-x)p\sim
\lambda$, and the eikonal line is accounted for by
$p_{J_\beta}=2$. The above result generalizes that for a single
quark state \cite{Berger:2002sv}.

For a pion, $I_{J_L} = I_{J_R}=2$, and the lowest infrared power
is $-1$, obtained when $p_{J_{L,R}}=1$, $p_{J_\beta}=2$,
$E^f_{J_L} = E^f_{J_R} = 1$, $E^f_\beta=0$,
$s_{J_\beta}=E^{b}_\beta$, and $s_{J_{L,R}} = E^b_{J_{L,R}}+1$.
The corresponding reduced diagram is shown in Fig. 2b, where the
final-state jet contains just the eikonal line and
longitudinally-polarized gluons, and the soft part contains an
anti-quark line. The jets going into the hard parts contain a
single quark and an arbitrary number of longitudinally polarized
gluons. We argue, however, that the actual infrared power is 0
when taking into account an additional numerator suppression, just
as in the leading-order diagram in Fig. 1. Observe that the
infrared power of an initial-state jet is always an half-integer,
which means the jet is proportional to at least one transverse
momentum, even after all integrals in the jet are carried out.
However, the only relevant transverse momenta available are those
of the soft lines. Therefore, this transverse momentum must be
counted as $\lambda$, not $\lambda^{1/2}$. With this extra
suppression, the actual infrared power of the reduced diagram in
Fig. 2b is 0. Factorizing out the transverse-momentum integral,
one gets $q(x)\sim (1-x)^2$ to all orders in perturbation theory.

The next lowest infrared power, 0, is obtained when $p_{J_R} =
p_{J_R} = 2$, $p_{J_\beta}=4$, all $E^f=0$ and $E^b_H=0$, and
$E^b_J = s_J$. The corresponding reduced diagram is shown in Fig.
2c, where the final-state jet has an anti-quark quantum number.
The soft-gluon radiation is not drawn because the initial and
final state jets are color-neutral; and thus when summing over all
final states, the soft factor reduces to 1, i.e., the soft
radiation must cancel. Because of the angular momentum
conservation, there is a numerator suppression factor proportional
to the transverse-momentum $k_\perp^2$ of the anti-quark jet, as
seen in Fig. 1a. The argument goes as follows. Consider
deep-inelastic scattering on a pion target. If the final state
consists of two collinear quark-antiquark jets plus an arbitrary
number of longitudinally polarized gluons going exactly along the
$n$ direction, the scattering amplitude vanishes identically. This
is because when the initial and final states are collinear and
with zero helicity (fermion helicity is conserved in gauge
theory), scattering with a physical photon of helicity $\pm 1$ is
forbidden by angular momentum conservation. Therefore, there must
be a relative transverse-momentum between the quark and anti-quark
jets, $k_\perp^2 \sim (1-x)Q^2$. The scattering amplitude is
proportional to $k_\perp$, and the parton distribution is
proportional to $k_\perp^2$. The actual infrared factor is then
$\lambda$. Accounting for the transverse-momentum integral, one
again gets a scaling law $(1-x)^2$ to all orders in perturbation
theory.


One can write down a factorization formula for the two reduced
diagrams shown in diagrams 2b and 2c, after factorizing the
collinear longitudinally polarized gluons from the hard parts and
the soft gluons from the initial state jets. For example, let us
consider the case when the anti-quark is soft (Fig. 2b). Going
through steps similar to those outlined in
\cite{CSS89,Berger:2002sv}, one finds
\begin{eqnarray}
  q_\pi(x\rightarrow 1) &=& H_L(p,\mu)H_R(p,\mu) \tilde J_{p,L}(p,\mu)
  \tilde J_{p,R}(p,\mu) \nonumber \\
   && \times \sum_{C_\beta,C_S} \prod_{m_L,m_R,l}\int
   \frac{d^nq_{L,m_L}}{(2\pi)^n}\frac{d^nq_{R,m_R}}{(2\pi)^n}
   \frac{d^n q_{l}}{(2\pi)^n} \nonumber \\
   && \times {\cal E}_L(p,\{q_{L,m_L}\})^{\{\gamma_{m_L}\}}
     {\cal E}_R(p,\{q_{R,m_R}\})^{\{\gamma_{m_R}\}} \nonumber \\
   && \times\int dy\int dz
   S^{(C_S)}\left(yp,\mu;\{q_{L,m_L}^{\gamma_{m_L}}\};\{q_{R,m_R}^{\gamma_{m_R}}\};\{
   q_l^\delta\}\right) \tilde J_\beta^{(C_\beta)}(zp,\mu; n;\{
   q_l^\delta\}) \nonumber \\
   && \times \delta^n\left(\sum_{m_L} q^\mu_{L,m_L} +
   \sum_{m_R} q^\mu_{R,m_R}  + \sum_l
   q_l^\mu\right)\delta(1-x-y-z) \ ,
\end{eqnarray}
where $H_{L,R}$ represent the hard contributions with collinear
longitudinally-polarized gluons factorized, $\tilde J_{L,R}$
represent the jet contributions with soft gluons and quarks
factorized. The factorized soft gluons are now represented by the
two eikonal lines ${\cal E}_{L,R}$. The soft loop momenta to be
integrated over connect the soft part and the eikonal lines, the
soft part and final-state jets. The summations are over all
possible cuts through final-state jet $C_\beta$ and soft part
$C_S$. The longitudinal momenta going through the final state cuts
are constrained by $1-x$. The soft function and jets can be
represented by non-perturbative QCD matrix elements.

One can define the gauge-invariant quark-eikonal cross section,
\begin{eqnarray}
   \sigma^{\rm eik}((1-x)p,\mu) = \frac{1}{N_c} \int \frac{dy^-}{2\pi} e^{i(1-x)p^+y^-}
       \sum_{n,a}\langle 0|\overline{\Psi}_a(y^-)|n\rangle
       \gamma^+\langle
       n|\Psi_a(0)|0\rangle\ ,
\end{eqnarray}
where $\Psi(y^-)$ is defined as
\begin{equation}
     \Psi(y^-) = \int^0_{-\infty} dy^+ U_n(\infty; y^-) U_P(0^+,y^-;
     y^+,y^-)\psi(y^+,y^-) \ ,
\end{equation}
with $U_n(\infty;y^-) = {\cal
P}\exp\left(-ig\int^\infty_{y^-}d\lambda n\cdot A(\lambda
n)\right)$, etc., and $P$ in $U_P$ is the pion momentum.
$\sigma^{\rm eik}$ has a similar factorization as the parton
distribution, except the pion jets are replaced by eikonal jets.
Therefore, $x\rightarrow 1$ quark distribution can be expressed
as,
\begin{equation}
    q(x) = H_L(p,\mu) H_R(p,\mu) J_L(p,\mu)  J_R(p,\mu) \sigma^{\rm eik}((1-x)p,\mu)\ .
\end{equation}
The entire $x$-dependence can now be found from the quark-eikonal
cross section. This result is very similar to the factorization of
the quark distribution in a single quark in the same limit
\cite{Korchemsky:1988si,Korchemsky:1992xv,Berger:2002sv}. It is
easy to check that at tree level, one reproduces the result from
Fig. 1.

Apart from the simple power behavior considered so far,
perturbative calculations also generate single and double
logarithmic dependence in $1-x$. The single logarithmic dependence
can be summed to yield a power correction $(1-x)^{A(\mu)}$, where
$A(\mu)$ is renormalization-scale $\mu$ dependent
\cite{Korchemsky:1988si}. The evolution in $\mu$ can be calculated
in perturbation theory, and the absolute magnitude of $A(\mu)$ can
only be determined through non-perturbative methods. Soft-gluon
radiations from charged lines produce large double logarithmic
corrections of type $\alpha_s^n \ln^m(1-x)$ with $m\le 2n-1$
\cite{BroLep80,Mueller:1981sg,Sterman:1986aj,Catani:1989ne,Manohar:2003vb}.
These corrections depend on a precise definition of the parton
distribution \cite{Sterman:1986aj} and infrared regularizations.
These issues will be discussed in a forth-coming publication.

To summarize, we have presented a factorization formula for the
valence quark distribution in the pion valid to the leading power
in $(1-x)$ and to all orders in perturbation theory. Through
infrared power counting, one obtains the power dependence in
$(1-x)$. The factorization theorem, however, allows one to
calculate the coefficients of the power law non-perturbatively. It
also allows one to re-sum large Sudakov double logarithms
associated with soft radiations.

We thank G. Sterman for useful discussions. X. J. is supported by
the U. S. Department of Energy via grant DE-FG02-93ER-40762, and
F. Y. by the RIKEN-BNL Research Center. J.P.M. was supported by
National Natural Science Foundation of P.R. China (NSFC). X. J. is
also supported by a grant from NSFC.

\end{document}